\documentclass[a4paper,
               ]{jacow}
\usepackage{pdfpages,multirow,ragged2e} %
\makeatletter%
	\ifboolexpr{bool{xetex}}
	 {\renewcommand{\Gin@extensions}{.pdf,%
	                    .png,.jpg,.bmp,.pict,.tif,.psd,.mac,.sga,.tga,.gif,%
	                    .eps,.ps,%
	                    }}{}
\makeatother

\ifboolexpr{bool{xetex} or bool{luatex}} 
 {}                                      
 {\usepackage[utf8]{inputenc}}           

\usepackage[USenglish]{babel}

\ifboolexpr{bool{jacowbiblatex}}%
 {%
  \addbibresource{jacow-test.bib}
  \addbibresource{biblatex-examples.bib}
 }{}
\begin{document}

\title{The SPARTA project: toward a demonstrator facility \\for multistage plasma acceleration}

\author{C. A. Lindstr{\o}m\thanks{c.a.lindstrom@fys.uio.no}, E. Adli, H. B. Anderson, P. Drobniak, D. Kalvik, F. Pe\~{n}a\textsuperscript{1}, K. N. Sjobak, \\ Department of Physics, University of Oslo, Oslo, Norway \\
		\textsuperscript{1}also at Ludwig-Maximilians-Universit{\"a}t M{\"u}nchen, Munich, Germany}
	
\maketitle

\begin{abstract}
   Plasma accelerators promise greatly reduced size and cost for future particle-accelerator facilities. However, several challenges remain to be solved; in particular that of coupling beams between plasma stages (i.e., \textit{staging}) without beam-quality degradation, and that of ensuring a stable acceleration process. In order to mature the technology, it is also key to identify an application that requires staging and high stability but is not overly challenging in other parameters such as energy efficiency, beam quality and repetition rate. The goal of the ERC-funded project SPARTA is to solve the staging and stability problems of plasma acceleration, and to combine the solutions into a medium-scale multistage plasma-accelerator facility for such an application: experiments in strong-field quantum electrodynamics. Here, we discuss the three main objectives of the SPARTA project: developing a nonlinear plasma lens for staging, developing self-stabilization mechanisms, and providing a conceptual design for a multistage demonstrator facility. 
\end{abstract}

\section{Introduction}

High-energy particle physics is headed for an impasse: a new energy-frontier particle collider is required, but the projected price tag of order ten billion euros is hindering progress toward its construction. Alternative accelerator technologies, which includes plasma acceleration~\cite{Veksler1956,Tajima1979,Chen1985,Ruth1985}, have therefore been identified as a potential long-term solution to this challenge: the Accelerator R\&D Roadmap of the 2020 European Strategy for Particle Physics~\cite{ESPP2020} cites that \textit{“Development and exploitation of laser/plasma acceleration techniques”} is one of \textit{“five key areas where an intensification of R\&D is required.”} However, the societal impact of more affordable high-energy particle beams goes beyond particle physics, ranging into diverse fields such as photon science (e.g., free-electron lasers \cite{Madey1971,McNeil2010}) and medical applications (e.g., cancer treatment and medical imaging \cite{ChaoChou2009}). 

Plasma accelerators \cite{Esarey2009,LindstromCorde2025} promise drastically to reduce the size and cost of future particle-accelerator facilities by providing significantly higher accelerating gradients \cite{Leemans2006,Hogan2005}, while simultaneously providing similar (or better) beam quality \cite{Lindstrom2021b,Lindstrom2024} and energy efficiency \cite{Litos2014,Pena2024} compared to radio-frequency (RF) accelerators. However, several challenges remain to be tackled before plasma accelerators can be used to replace or augment RF accelerators for high-energy applications. Two of the core challenges are: (i) connecting multiple plasma-accelerator stages (i.e., staging) \cite{Steinke2016,Lindstrom2021a} without loss of beam quality; and (ii) stabilization of the acceleration process, to be resistant to imperfections and instabilities \cite{Whittum1991,Lebedev2017}. While there are several other important challenges, including positron acceleration \cite{Cao2024}, high-repetition-rate operation \cite{Darcy2022}, and preservation of ultra-low emittance, for high-energy applications (and in particular colliders) these issues are all secondary to demonstrating a stable and energy-scalable scheme for accelerating electrons.

Another challenge for high-energy plasma acceleration, even after a solution for staging and stability has been identified, is that the ultimate application, a plasma-based linear collider \cite{Seryi2009,Foster2023}, requires extreme performance and reliability in all aspects simultaneously---a prohibitively large step up from state-of-the-art plasma-accelerator experiments. In order to mature a disruptive technology like plasma acceleration more gradually, a simpler application is required to make near-term progress. While much progress has been made toward a plasma-based free-electron laser \cite{Wang2021,Pompili2022}, which requires high beam brightness, this application does not necessarily align with the need for staging and stability. Experiments in strong-field quantum electrodynamics \cite{Bula1996,Cole2018} (SFQED), on the other hand, are highly aligned with these needs, making it an ideal first application. Typically, SFQED is probed by colliding a high-intensity (>\SI{e20}{W/cm^2}) laser pulse with a high-energy (>\SI{10}{GeV}) electron beam---in the rest frame of the electrons, the electromagnetic fields of the laser pulse appear relativistically contracted, and hence boosted by the electron's Lorentz factor, approaching or exceeding the Schwinger field~\cite{Schwinger1951} (\SI{1.3e18}{V/m} or \SI{4.4e9}{T}). For this application, stable and high-energy electron beams are required, but only modest beam quality, energy efficiency and repetition rate.

\begin{figure*}[t]
   \centering
   \includegraphics*[width=0.92\textwidth]{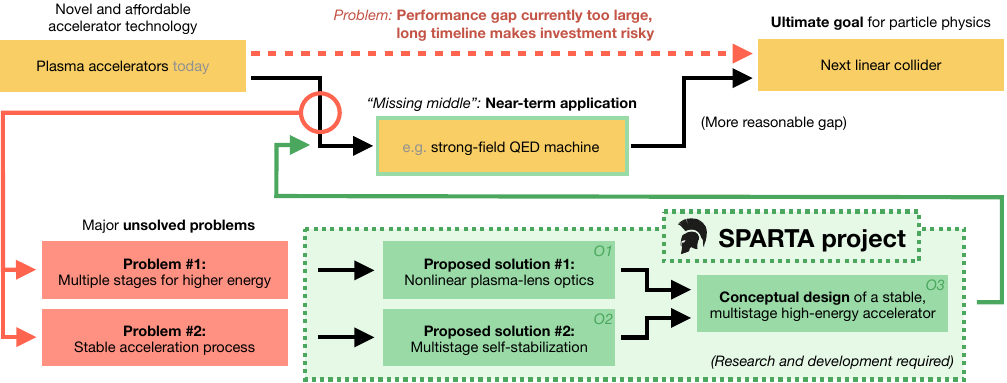}
   \caption{Flow chart describing the context and motivations (yellow boxes), challenges to be solved (red boxes) and proposed solutions (green boxes) for the SPARTA project.}
   \label{fig:flow-chart}
\end{figure*}

The SPARTA project (Staging of Plasma Accelerators for Realizing Timely Applications), funded by the ERC \cite{SPARTA} for the period 2024--2029, aims to tackle the two above-mentioned core challenges in plasma acceleration (i.e., staging and stability) and to deliver blueprints for a medium-term multistage demonstrator facility. This is illustrated by the flow chart in Fig.~\ref{fig:flow-chart}. In this paper, we break down the three main objectives of the project: (1) the development of a novel \textit{nonlinear plasma lens}, which promises to solve the staging problem; (2) investigating \textit{self-stabilization mechanisms} in the longitudinal and transverse phase space, which promise to solve the stability problem; and (3) the conceptual design, using a newly developed start-to-end simulation framework, of a multistage plasma accelerator---a demonstrator facility with the capability of performing beyond-state-of-the-art SFQED experiments.

\section{Objective 1: develop \\ a nonlinear plasma lens}

\begin{figure}[b]
   \centering
   \includegraphics*[width=\linewidth]{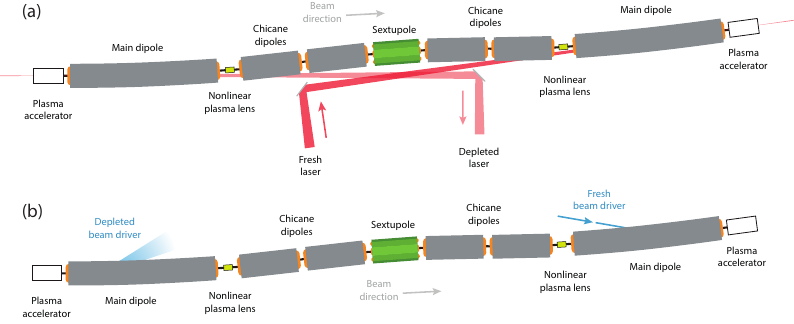}
   \caption{Staging optics lattice based on nonlinear plasma lenses, placed between two plasma accelerators driven either by laser drivers (a) or beam drivers (b).}
   \label{fig:staging-lattice}
\end{figure}

The challenge of staging is multi-faceted \cite{Lindstrom2021b}: the transport of highly diverging and large-energy-spread beams between plasma-accelerator stages results in chromaticity (energy-dependent focusing) and consequently beam-quality degradation; refocusing and chromaticity-correcting high-energy beams typically takes up significant space; and, lastly, the wakefield drivers need to be in- and out-coupled.

Plasma lenses, which can focus beams much more strongly than magnetic quadrupoles as well as focus simultaneously in both planes, promise to enable compact staging optics. Both passive plasma lenses \cite{Chen1987,Thaury2015} (i.e., very short plasma accelerators) and active plasma lenses \cite{vanTilborg2015,Lindstrom2018} (i.e., gas-filled tubes with a high-current discharge running parallel to the beam) can be used. Active plasma lenses typically require a less complex setup and were already successfully employed in the first proof-of-principle staging laser-wakefield-driven experiment at LBNL \cite{Steinke2016}. While very compact, this experiment still suffered from chromaticity and used plasma mirrors for in-coupling the drivers (not applicable to beam-driven plasma accelerators). 

To solve these remaining issues, a novel staging lattice is proposed, as illustrated in Fig.~\ref{fig:staging-lattice}. This lattice is based on local chromaticity correction \cite{Raimondi2001}, i.e., the combination of dipole dispersion and nonlinear focusing fields (typically provided by sextupoles) at the location of the strongest focusing elements. Here, however, the idea is to integrate a sextupole-like field into the plasma lens itself, resulting in a \textit{nonlinear plasma lens}---a beamline element that does not yet exist. Objective \#1 of the SPARTA project is therefore to experimentally develop such a nonlinear plasma lens.

This development work has recently started, as reported in Refs.~\cite{Drobniak2025a} and \cite{Drobniak2025b}. First experiments, with a prototype lens using an external magnetic dipole field across the discharge capillary to form the desired transverse magnetic-field distribution, were performed at the CLEAR user facility \cite{Gamba2018} at CERN in Sep.~2024, with more experiments planned in 2025. In parallel, hydrodynamic simulations using COMSOL are being performed. Once a working prototype lens has been made, a first proof-of-principle demonstration of the plasma-lens-based local chromaticity correction will be performed in the form of a novel \textit{achromatic spectrometer} for diagnosing high-divergence and high-energy-spread beams from a laser-wakefield accelerator (see Ref.~\cite{Pena2025}).

\section{Objective 2: develop self-stabilization mechanisms}

\begin{figure*}[t]
   \centering
   \includegraphics*[width=0.93\textwidth]{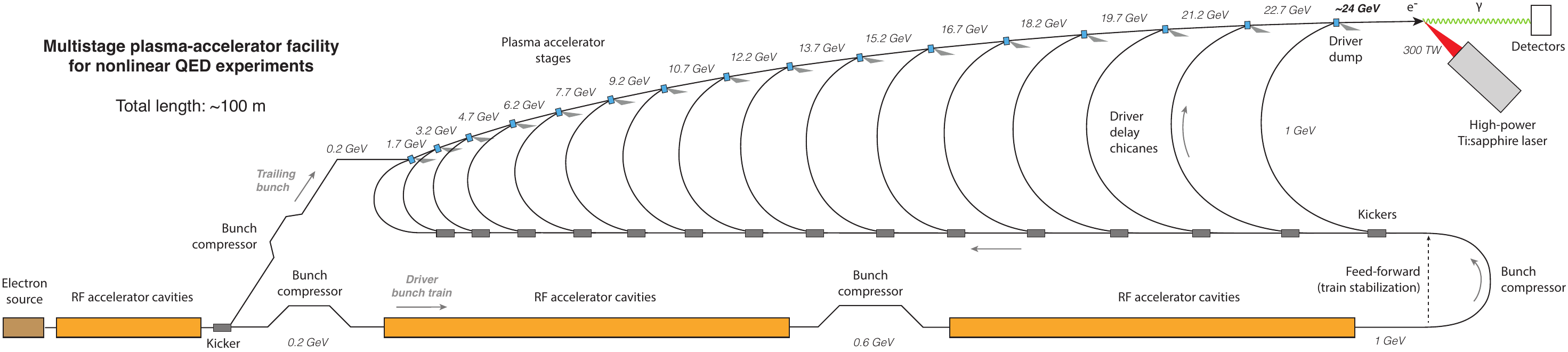}
   \caption{Schematic layout of a possible beam-driven multistage plasma accelerator. The drive-beam distribution concept indicated here is only a preliminary sketch and will likely be significantly more compact and simplified in the final design.}
   \label{fig:demo-facility}
\end{figure*}

Plasma-accelerator experiments often suffer from poor stability. This results from a combination of (i) the very small length and time scales (\SI{}{\micro\m}--mm and fs--ps, respectively) making even small absolute jitters relatively large, and (ii) intrinsic instabilities stemming from beam--plasma resonances. Attention has recently turned to improving the stability, both using active stabilization \cite{Maier2020}, which involves rapid measurement and adjustment of parameters, as well as passive stabilization (or \textit{self-stabilization}). While the former will always be required to a certain degree, it adds complexity and cost, whereas the latter tends to greatly simplify operation. A first step toward passive stabilization was recently made in a laser-wakefield experiment at DESY \cite{Winkler2025}, where the energy was stabilized using a magnetic chicane and an RF cavity. Objective \#2 of the SPARTA project is to investigate and develop self-stabilization mechanisms in both the longitudinal and transverse phase space, in particular utilizing multistage dynamics, as moving in and out of stages enables the employment of (negative) feedback loops that can automatically correct or damp errors.

A powerful multistage self-stabilization mechanism in the longitudinal phase space was recently proposed \cite{Lindstrom2021c}, making use of longitudinal dispersion ($R_{56}$) between stages---this allows particles that were accelerated too much in one stage to move longitudinally forward, such that the accelerating field experienced in the next stage is smaller, and vice versa. Performing large-scale start-to-end simulations with the newly-developed ABEL simulation framework \cite{Chen2025}, the SPARTA project will investigate the practical applicability of this mechanism, including how it is affected by the specific staging optics, synchrotron radiation and more.

Transverse instabilities such as hosing or beam-breakup \cite{Whittum1991,Lebedev2017} must be fully suppressed in order to stably operate an efficient and high-gain plasma accelerator. Mitigation strategies include, in particular, controlled use of ion motion \cite{Mehrling2018}. The SPARTA project will perform experimental investigations into transverse instability \cite{Finnerud2024} and ion motion \cite{Kalvik2025} at FACET-II, SLAC \cite{Yakimenko2019} and FLASHForward, DESY \cite{Darcy2019}. The possibility of multistage transverse self-stabilization will also be investigated with numerical simulations.

\section{Objective 3: design a multistage plasma-accelerator facility}

Building on the solutions identified in Objectives \#1 and \#2, Objective \#3 is to design a multistage plasma-accelerator facility, at a level of detail and credibility sufficient for the potentially large financial investment required (€10--100M scale). This work will proceed in two steps, described below.

The first step is to design a proof-of-principle staging experiment testing the staging lattice illustrated in Fig.~\ref{fig:staging-lattice}. This could in principle be performed between two plasma-accelerator stages, but does not have to---it may be beneficial to transport and diagnose more controlled beams from an RF accelerator to emulate plasma-accelerated beams. Designing such an experiment requires considering many details such as alignment tolerances, magnet specifications, vacuum systems, beam diagnostics and beam dumps. 

Secondly, building on the detailed knowledge gained from designing a two-stage experiment, the conceptual design of a medium-scale SFQED machine (as illustrated in Fig.~\ref{fig:demo-facility}) will be performed. This work will be done using the ABEL framework, combining particle-in-cell simulations (e.g., with HiPACE++), tracking simulations (e.g., with ImpactX) and SFQED simulations (e.g., with Ptarmigan). ABEL also allows cost modelling and optimization, which will be performed within the specified restrictions that exist at e.g.~SLAC or DESY; sites where such a machine may feasibly be implemented. Currently, the aim is to deliver \SI{\sim50}{GeV} electron bunches at reasonable charge (0.1--\SI{1}{nC}), emittance (\SI{\sim10}{\milli\m\milli\radian}) and repetition rate (1--\SI{10}{Hz}) using around 10 plasma stages and a total length of \SI{\sim100}{m}.

\section{Conclusion}
The SPARTA project is making steady progress, both experimentally and theoretically, toward providing stable high-energy electron beams from multistage plasma accelerators. If the ongoing experimental demonstration of nonlinear plasma lenses is successful, and a credible design for a medium-scale multistage demonstrator facility can be developed, this will pave the way toward applications such as a SFQED experiment and, in the longer term, a plasma-based collider.

\section{ACKNOWLEDGEMENTS}
This work is funded by the European Research Council (ERC Grant Agreement No. 101116161).

%
%
\ifboolexpr{bool{jacowbiblatex}}%
	{\printbibliography}%

\begin{thebibliography}{99}   

    \bibitem{Veksler1956}
    V. I. Veksler, 
    ``Coherent principle of acceleration of charged particles", 
    in Proc. CERN Symposium on High Energy Accelerators and Pion Physics (CERN, 1956), pp. 80--83.
    
    \bibitem{Tajima1979}
    T. Tajima and J. M. Dawson,
    ``Laser electron accelerator",
    \textit{Phys. Rev. Lett.} \textbf{43}, 267 (1979).
    
    \bibitem{Chen1985}
    P. Chen, J. M. Dawson, R. W. Huff, and T. Katsouleas, 
    ``Acceleration of electrons by the interaction of a bunched electron beam with a plasma",
    \textit{Phys. Rev. Lett.} \textbf{54}, 693 (1985).
    
    \bibitem{Ruth1985}
    R. D. Ruth, A. W. Chao, P. L. Morton, and P. B. Wilson,
    ``A plasma wake field accelerator",
    \textit{Part. Accel.} \textbf{17}, 171 (1985).

    \bibitem{ESPP2020}
    N. Mounet (ed.), 
    “European Strategy for Particle Physics - Accelerator R\&D Roadmap”, 
    CERN-2022-001 (CERN, 2022). 

    \bibitem{Madey1971}
    J. M. J. Madey,
    ``Stimulated emission of bremsstrahlung in a periodic magnetic field",
    \textit{J. Appl. Phys.} \textbf{42}, 1906 (1971).

    \bibitem{McNeil2010}
    B. W. J. McNeil and N. R. Thompson, 
    “X-ray free-electron lasers”, 
    \textit{Nat. Photon.} \textbf{4}, 814 (2010).

    \bibitem{ChaoChou2009}
    A. W. Chao and W. Chou (eds.), 
    Reviews of Accelerator Science and Technology, Volume 2: Medical Applications of Accelerators (World Scientific, Singapore, 2009).

    \bibitem{Esarey2009}
    E. Esarey, C. B. Schroeder and W. Leemans, 
    “Physics of laser-driven plasma-based electron accelerators”, 
    \textit{Rev. Mod. Phys.} \textbf{81}, 1229 (2009).

    \bibitem{LindstromCorde2025}
    C. A. Lindstr{\o}m, S. Corde \textit{et al.},
    “Beam-driven plasma wakefield acceleration”, 
    preprint at arXiv:2504.05558 (2025).
    
    \bibitem{Hogan2005}
    M. J. Hogan \textit{et al.},
    ``Multi-GeV energy gain in a plasma-wakefield accelerator",
    \textit{Phys. Rev. Lett.} \textbf{95}, 054802 (2005).

    \bibitem{Leemans2006}
    W. P. Leemans \textit{et al.}, 
    ``GeV electron beams from a centimetre-scale accelerator",
    \textit{Nat. Phys.} \textbf{2}, 696 (2006).
    
    \bibitem{Lindstrom2021b}
    C. A. Lindstr{\o}m \textit{et al.},
    “Energy-spread preservation and high eﬃciency in a plasma-wakefield accelerator”,
    \textit{Phys. Rev. Lett.} \textbf{126}, 014801 (2021).

    \bibitem{Lindstrom2024}
    C. A. Lindstr{\o}m \textit{et al.},
    “Emittance preservation in a plasma-wakefield accelerator”,
    \textit{Nat. Commun.} \textbf{15}, 6097 (2024).

    \bibitem{Litos2014}
    M. Litos \textit{et al.},
    ``High-efficiency acceleration of an electron beam in a plasma wakefield accelerator",
    \textit{Nature} \textbf{515}, 92 (2014).

    \bibitem{Pena2024}
    F. Pe{\~n}a \textit{et al.},
    ``Energy depletion and re-acceleration of driver electrons in a plasma-wakefield accelerator",
    \textit{Phys. Rev. Research} \textbf{6}, 043090 (2024).
    
    \bibitem{Steinke2016}
    S. Steinke \textit{et al.},
    “Multistage coupling of independent laser-plasma accelerators”, 
    \textit{Nature} \textbf{530}, 190 (2016).

    \bibitem{Lindstrom2021a}
    C. A. Lindstr{\o}m, 
    “Staging of plasma-wakefield accelerators”, 
    \textit{Phys. Rev. Accel. Beams} \textbf{24}, 014801 (2021).

    \bibitem{Whittum1991}
    D. H. Whittum \textit{et al.},
    “Electron-hose instability in the ion-focused regime”, 
    \textit{Phys. Rev. Lett.} \textbf{67}, 991 (1991).

    \bibitem{Lebedev2017}
    V. Lebedev, A. Burov and S. Nagaitsev,
    “Eﬃciency versus instability in plasma accelerators”, 
    \textit{Phys. Rev. Accel. Beams} \textbf{20}, 121301 (2017).

    \bibitem{Cao2024}
    G. J. Cao \textit{et al.},
    ``Positron acceleration in plasma wakefields",
    \textit{Phys. Rev. Accel. Beams} \textbf{27}, 034801 (2024).

    \bibitem{Darcy2022}
    R. D'Arcy \textit{et al.},
    “Recovery time of a plasma-wakefield accelerator”, 
    \textit{Nature} \textbf{603}, 58 (2022).

    \bibitem{Seryi2009}
    A. Seryi \textit{et al.},
   ``A Concept of Plasma Wake Field Acceleration Linear Collider (PWFA-LC)",
   in \emph{Proc. PAC’09}, Vancouver, Canada, May 2009, paper WE6PFP081, pp. 2688--2690.

    \bibitem{Foster2023}
    B. Foster, R. D'Arcy and C. A. Lindstr{\o}m, 
    “A hybrid, asymmetric, linear Higgs factory based on plasma-wakefield and radio-frequency acceleration”, 
    \textit{New J. Phys.} \textbf{25}, 093037 (2023).
    
    \bibitem{Wang2021}
    W. Wang \textit{et al.},
    ``Free-electron lasing at 27 nanometres based on a laser wakefield accelerator",
    \textit{Nature} \textbf{595}, 516 (2021);

    \bibitem{Pompili2022}
    R. Pompili \textit{et al.},
    ``Free-electron lasing with compact beam-driven plasma wakefield accelerator", 
    \textit{Nature} \textbf{605}, 659 (2022).

    \bibitem{Bula1996}
    C. Bula \textit{et al.},
    ``Observation of nonlinear effects in Compton scattering", 
    \textit{Phys. Rev. Lett.} \textbf{76}, 3116 (1996).

    \bibitem{Cole2018}
    J. Cole \textit{et al.},
    ``Experimental evidence of radiation reaction in the collision of a high-intensity laser pulse with a laser-wakefield accelerated electron beam",
    \textit{Phys. Rev. X} \textbf{8}, 011020 (2018);
    
    \bibitem{Schwinger1951}
    J. Schwinger, 
    ``On gauge invariance and vacuum polarization", 
    \textit{Phys. Rev.} \textbf{82}, 664 (1951).

    \bibitem{SPARTA}
    European Commission, 
    ``Staging of plasma accelerators for realizing timely applications", 2023. \\
    \url{https://doi.org/10.3030/101116161}.

    \bibitem{Chen1987}
    P. Chen, 
    “A possible final focusing mechanism for linear colliders”, 
    \textit{Part. Accel.} \textbf{20}, 171 (1987).
    
    \bibitem{Thaury2015}
    C. Thaury \textit{et al.},
    “Demonstration of relativistic electron beam focusing by a laser-plasma lens”, 
    \textit{Nat. Commun.} \textbf{6}, 6860 (2015).

    \bibitem{vanTilborg2015}
    J. van Tilborg \textit{et al.},
    “Active plasma lensing for relativistic laser-plasma-accelerated electron beams”, 
    \textit{Phys. Rev. Lett.} \textbf{115}, 184802 (2015).

    \bibitem{Lindstrom2018}
    C. A. Lindstr{\o}m \textit{et al.},
    “Emittance preservation in an aberration-free active plasma lens”, 
    \textit{Phys. Rev. Lett.} \textbf{121}, 194801 (2018).
    
    \bibitem{Raimondi2001}
    P. Raimondi and A. Seryi, 
    “Novel final focus design for future linear colliders”, 
    \textit{Phys. Rev. Lett.} \textbf{86}, 3779 (2001).

    \bibitem{Drobniak2025a}
    P. Drobniak \textit{et al.},
    ``Development of a nonlinear plasma lens for achromatic beam transport", 
    \textit{Nucl. Instrum. Methods Phys. Res. A} \textbf{1072}, 170223 (2025).

    \bibitem{Drobniak2025b}
    P. Drobniak \textit{et al.},
    ``Preliminary results from the CLEAR nonlinear plasma lens experiment", 
     presented at the 16th International Particle Accelerator Conf. (IPAC’25), Taipei, Taiwan, June 2025 paper MOPS027, this conference.
    
    \bibitem{Gamba2018}
    D. Gamba \textit{et al.},
    ``The CLEAR user facility at CERN", 
    \textit{Nucl. Instrum. Methods Phys. Res. A} \textbf{909}, 480 (2018).

    \bibitem{Pena2025}
    F. Pe{\~n}a \textit{et al.},
    ``Development of an achromatic spectrometer for a laser-wakefield-accelerator experiment", 
     presented at the 16th International Particle Accelerator Conf. (IPAC’25), Taipei, Taiwan, June 2025 paper TUPM096, this conference.

    \bibitem{Maier2020}
    A. R. Maier \textit{et al.},
    ``Decoding sources of energy variability in a laser-plasma accelerator",
    \textit{Phys. Rev. X} \textbf{10}, 031039 (2020).
    
    \bibitem{Winkler2025}
    P. Winkler \textit{et al.},
    ``Active energy compression of a laser-plasma electron beam",
    \textit{Nature} \textbf{640}, 907 (2025).

    \bibitem{Lindstrom2021c}
     C. A. Lindstr{\o}m,
    ``Self-correcting longitudinal phase space in a multistage plasma accelerator",
    arXiv:2104.14460 (2021).

    \bibitem{Chen2025}
    J. B. B. Chen \textit{et al.},
    ``ABEL: The adaptable beginning-to-end linac simulation framework", 
    presented at the 16th International Particle Accelerator Conf. (IPAC’25), Taipei, Taiwan, June 2025 paper TUPS012, this conference.

    \bibitem{Mehrling2018}
    T. J. Mehrling \textit{et al.},
    ``Suppression of Beam Hosing in Plasma Accelerators with Ion Motion",
    \textit{Phys. Rev. Lett.} \textbf{121}, 264802 (2018).
        
    \bibitem{Finnerud2024}
     O. G. Finnerud \textit{et al.},
    ``The E302 instability-versus-efficiency experiment at FACET-II",
    arXiv:2402.10325 (2024).

    \bibitem{Kalvik2025}
    D. Kalvik \textit{et al.},
    ``Ion-motion simulations of a plasma-wakefield experiment at FLASHForward",
    presented at the 16th International Particle Accelerator Conf. (IPAC’25), Taipei, Taiwan, June 2025 paper TUPS013, this conference.

    \bibitem{Yakimenko2019}
    V. Yakimenko \textit{et al.},
    ``FACET-II facility for advanced accelerator experimental tests",
    \textit{Phys. Rev. Accel. Beams} \textbf{22}, 101301 (2019).
    
    \bibitem{Darcy2019}
    R. D'Arcy \textit{et al.},
    ``FLASHForward: plasma wakefield accelerator science for high-average-power applications",
    \textit{Phil. Trans. R. Soc. A.} \textbf{377}, 20180392 (2019).
    
	\end{thebibliography}
	{%

} 
%
%


\end{document}